\documentclass[aps,prb,twocolumn,showpacs,superscriptaddress]{revtex4}

\usepackage{amsmath}
\usepackage{amsthm}
\usepackage{amssymb}
\usepackage{graphicx}
\usepackage{tabularx}
\usepackage[squaren]{SIunits}

\def\be{\begin{equation}}
\def\ee{\end{equation}}
\def\bea{\begin{eqnarray}}
\def\eea{\end{eqnarray}}
\def\ba{\begin{array}}
\def\ea{\end{array}}

\begin{document}


\title{Kondo effect with diverging hybridization: possible realization in \\graphene with vacancies}

\author{Andrew K. Mitchell}
\affiliation{Institut f\"ur Theoretische Physik, Universit\"at zu K\"oln, Z\"ulpicher Stra\ss e 77, 50937 K\"oln, Germany}
\affiliation{Department of Chemistry, Physical and Theoretical Chemistry,
Oxford University, South Parks Road, Oxford OX1 3QZ, United Kingdom}
\author{Lars Fritz}
\affiliation{Institut f\"ur Theoretische Physik, Universit\"at zu K\"oln, Z\"ulpicher Stra\ss e 77, 50937 K\"oln, Germany}

\date{\today}


\begin{abstract}
We investigate Kondo physics in a host with a strongly diverging density of states. This study is motivated by a recent work on vacancies in the graphene honeycomb lattice, whose density of states is enhanced at low energies due to potential scattering.
The generalized quantum impurity model describing the vacancy is shown to support a spin-$\tfrac{1}{2}$ (doublet) Kondo phase. The special role played by a diverging host density of states is examined in detail, with distinctive signatures associated with the powerlaw Kondo effect shown to appear in thermodynamic quantities and the scattering t matrix, with a strongly enhanced Kondo temperature. Although the effective Kondo model supports a novel stable phase characterized by strong renormalized particle-hole asymmetry, we find that this phase cannot in fact be accessed in the full Anderson model. In the more realistic case where the divergence in the host density of states is cut off at low energies, a crossover is generated between pristine powerlaw Kondo physics and a regular Kondo strong coupling state. 
\end{abstract}

\pacs{73.22.Pr, 72.15.Qm, 72.10.Fk}

\maketitle


In standard metals, local magnetic moments are screened by conduction electrons at low temperatures, which together form a quantum many-body spin-singlet by the Kondo effect.\cite{hewson} 
Precise details of the host band structure do not affect the universal physics --- provided\cite{hewson,nozieres,WF} the density of states is essentially flat at low energies. The Kondo effect has thus become a paradigm for strong correlations in condensed matter science. 

However, the fate of local magnetic moments coupled to low-energy degrees of freedom in unconventional materials is much more subtle, and has been the subject of enduring investigation over the last twenty years. Fascinating variants of the classic Kondo paradigm have been variously sought in e.g. high-temperature superconductors\cite{WF,GBI,VF04,FV04} and spin liquids\cite{kolezhuk06,florens06} containing magnetic impurities. More recently, graphene has been the focus of considerable attention, due to its unusual density of states.\cite{novo1,novo2,RMPNeto} As a consequence, magnetic impurities in graphene may exhibit unusual Kondo physics\cite{zhuang09,cornaglia09,uchoa08,uchoa09,berakdar10,VFB10}, and even local quantum phase transitions.\cite{WF,GBI,VF04,FV04,VFB10} 

It was also appreciated that \emph{defects} in graphene --- such as carbon vacancies and induced reconstructions in the otherwise perfect honeycomb lattice --- can give rise to
interesting new physics.\cite{balseiro12,yndurain12,pruschke11,kondo_defect_graphene,cazalilla12} 
In particular, evidence of the Kondo effect has been observed
experimentally in irradiated graphene
samples which host such vacancies.\cite{grigorieva12,kawakami12,fuhrer10}  
It was argued recently by Cazalilla {\it et al.} in
Ref.~\onlinecite{cazalilla12} that structural corrugations around 
vacancies allow for a hybridization between the $\sigma$- and $\pi$-orbitals,
which leads to local magnetic moment formation (an effect absent in flat
graphene). Potential scattering from the defect induces a 
greatly enhanced density of conduction electron states coupling to
these local moments.\cite{cazalilla12} This would give rise to very high Kondo
temperatures --- as in fact observed in certain experiments.\cite{fuhrer10}
For a thorough recent review of Kondo physics in graphene, see Ref.~\onlinecite{FV12}. 


In this paper we study the effective model for isolated
vacancies in graphene introduced in Ref.~\onlinecite{cazalilla12}. A particular limit of the full model is explicitly considered, in which interactions between the $\sigma$ and $\pi$ systems are small (although \emph{local} interactions may be strong). In this case, we show that the unusual graphene density of states near the vacancy leads to a variant of the spin-$\tfrac{1}{2}$ pseudogap Kondo problem.\cite{WF,GBI,VF04,FV04,loganpt,lmapg,BGLlma} The divergent hybridization represents a limiting case, and exotic quantum impurity physics results. For example, the residual impurity entropy of $S_{\text{imp}}=-k_B\ln(4)$ is the lowest possible for this type of system. 

We go beyond the previous perturbative analysis using the full 
numerical renormalization group (NRG) to calculate various physical
quantities exactly (for a recent review of the technique, 
see Ref.~\onlinecite{nrg:rev}). Numerical results are supported
wherever possible by analytical arguments. We also consider the eventual low-temperature crossover to a regular Kondo strong coupling state, induced when the divergence in the host density of states is cut off, as might be expected in real systems. We note that the full model, including also $\sigma$-$\pi$ electronic interactions, is more complex, and can support for example a pseudogapped local moment phase. The full phase diagram of the model on increasing $\sigma$-$\pi$ interactions will be to subject of a forthcoming publication.\cite{graphene2} 

The main results of this paper are as follows: \\
(i) The phase diagram of the effective Kondo model is presented and
discussed. The modified powerlaw density of states leads to large Kondo temperatures, and supports two distinct strong coupling phases, corresponding to runaway
RG flow of either the Kondo exchange coupling or potential scattering. 
Thermodynamic and dynamical quantities are analyzed analytically at
the various stable fixed points, while the full crossover between
fixed points is calculated numerically. \\
(ii) We find that the particle-hole asymmetric local moment phase of
the simplified Kondo model is \emph{not} accessible in the
underlying full Anderson model, and hence will not in practice be 
realized in experiments.\\
(iii) Crossover from pristine powerlaw Kondo physics to a standard Kondo strong coupling state arises when the diverging host density of states is cut off at low energies, as might be expected in real systems. However, physical properties may still be controlled by the powerlaw Kondo effect over an extended temperature/energy window. 

The organization of the manuscript is as follows. 
In Sec.~\ref{sec:model} we introduce a generalized Anderson impurity model 
describing a single reconstructed graphene vacancy, following Ref.~\onlinecite{cazalilla12}. In the limit of negligible $\sigma$-$\pi$ band interactions, an effective low-energy spin-$\tfrac{1}{2}$ Kondo model is derived and studied explicitly in Sec.~\ref{sec:effkondo}. Thermodynamic and dynamical quantities are calculated in each phase of
the model, and the respective phase diagram is analyzed in detail for the case of a diverging host density of states. Physical behavior in the vicinity of the stable strong coupling fixed points is analytically understood and interpreted in terms of an effective resonant level or pure potential scatterer. 
The evolution of strong coupling scales is extracted in each case.  
We also examine the quantum phase transition separating Kondo screened
and asymmetric local moment states.

In Sec.~\ref{sec:anderson} we analyze the more physical Andersonian model, and
find that the potential scattering phase of the Kondo model is not in fact
accessible.  We go on to consider the effect of cutting off the powerlaw divergence in the conduction electron density of states at low energies in Sec.~\ref{sec:genmodel}. The resulting crossover to conventional Kondo physics is examined.

We conclude with a discussion of our results and their relevance to experiments on irradiated graphene in Sec.~\ref{sec:conclusions}. More technical parts of the 
analysis are relegated to the appendix.


\section{Model and observables}\label{sec:model}

Cazalilla \emph{et al.} derived an effective model in Ref.~\onlinecite{cazalilla12}, describing a single vacancy in graphene. It is formulated in terms of a single `dangling' quantum orbital of the $\sigma$ band localized at the defect site (denoted $d_{\nu}$), which hybridizes with conduction electrons of the $\pi$ band --- an effect precluded in flat graphene by the symmetry of the orbitals, but allowed near a defect due to structural distortions. The Coulomb interaction gives rise to a strong onsite electronic correlation at the $d$-level. A capacitive interaction between charge density in the $d$-level and nearby $\pi$ orbitals may also play a role, as might a Hund's rule coupling of their spin density.\cite{cazalilla12}

We thus take a generalized Anderson Hamiltonian
$H = H^{0}_{\pi} + H_{d}+ H_{\rm{hyb}} + H_{d\pi}$, where $H^{0}_{\pi} =
\sum_{p,\nu} \epsilon^{\phantom{\dagger}}_{p} \pi^\dagger_{p\nu}
\pi^{\phantom{\dagger}}_{p\nu}$ describes the free $\pi$ band conduction electrons of spin $\nu=\uparrow/\downarrow$ in terms of diagonal eigenmodes $\pi_{p\nu}$ (these operators already account for the presence of the vacancy), and
\begin{eqnarray}\label{fullH}
H_{d} &=& \epsilon^{\phantom{\dagger}}_d \hat{n}^{\phantom{\dagger}}_d + U^{\phantom{\dagger}}_{dd} \hat{n}_{d}^\uparrow \hat{n}_d^\downarrow \nonumber \\  
H_{\rm{hyb}} &=&  \sum_{p,\nu} \left(g_p \pi^{\dagger}_{p\nu} d^{\phantom{\dagger}}_\nu  +{\rm{H.c.}} \right) \nonumber\\
H_{d\pi} &=& \sum_{p,p',\nu} \left ( U^{\phantom{\dagger}}_{d\pi} \hat{n}^{\phantom{\dagger}}_d \pi^\dagger_{p,\nu}\pi^{\phantom{\dagger}}_{p',\nu} \right )  - J_{H} {\bf{S}}_d \cdot
{\bf{S}}_{\pi}   \;.  
\end{eqnarray}
Here, $\hat{n}_d=\sum_{\nu} \hat{n}_d^{\nu}=\sum_{\nu} d^{\dagger}_\nu
d^{\phantom{\dagger}}_{\nu}$ is the total number operator for the
$d$-level, and 
the spin densities of the $d$-level and $\pi$ band at the vacancy are  
given by ${\bf{S}}_d=\frac{1}{2} d^\dagger_{\nu} \vec{\sigma}_{\nu \nu'} 
d^{\phantom{\dagger}}_{\nu'}$ and ${\bf{S}}_{\pi}=\frac{1}{2} \sum_{p,p'}
\pi^\dagger_{p,\nu} \vec{\sigma}_{\nu \nu'} \pi^{\phantom{\dagger}}_{p',\nu'}$, 
interact via the Hund's coupling $J_{H}$.  
The $d$-level hybridizes with an \emph{effective} $\pi$ orbital, which from $H_{\text{hyb}}$ follows as $\tilde{\pi}_{\nu} = \frac{1}{g} \sum_p g_p\pi_{p\nu}$, with $g^2=\sum_p|g_p|^2$. The local density of states seen by the $d$-level is thus,
\begin{equation}\label{dosdef}
\begin{split}
\rho(\omega) &= \frac{1}{g^2}\sum_{p} |g_p|^2 \delta(\omega-\epsilon_{p})= -\frac{1}{\pi} \text{Im}~G_{\text{loc}}^{(0)}(\omega) \;,
\end{split}
\end{equation}
where $G_{\text{loc}}^{(0)}(\omega)$ is the retarded free electronic Green function of the local $\tilde{\pi}$ orbital.

Importantly, potential scattering from the vacancy gives rise to strongly modified properties of the free $\pi$-band conduction electrons,\cite{cazalilla12} which has been absorbed into the definition of the $\pi_p$ eigenmodes in Eq.~(\ref{fullH}), and thus enters implicitly through $\epsilon_p$ in $H_{\pi}^0$. 
Within a tight-binding nearest-neighbour hopping approximation, the density of states takes an unusual divergent form at low energies,\cite{cazalilla12}
\begin{eqnarray}\label{eq:DOS}
\rho (\omega) \sim \frac{1}{\left | \tfrac{\omega}{\Lambda} \right |
  \ln^2 \left | \tfrac{\omega}{\Lambda} \right |}~\;,
\end{eqnarray}
where $\Lambda$ is the energy scale below which the complicated band structure of the full defective graphene system becomes well-approximated by Eq.~(\ref{eq:DOS}).

The hybridization function for the $d$-level, $\Delta(\omega)=g^2
G^{(0)}_{\text{loc}} (\omega)$, contains all the information about the
non-interacting bath of conduction electrons and, together with
$H_{d}$ and $H_{d\pi}$, completely specify the underlying quantum impurity
problem. $G^{(0)}_{\text{loc}} (\omega)$ itself can be obtained via Eqs.~(\ref{dosdef}) and (\ref{eq:DOS}), provided the local density of states of the effective $\tilde{\pi}$ orbital is normalizable. Thus we introduce a high-energy band cutoff $D_0$, defining 
$\rho (\omega) \propto \theta\left(D_0-|\omega|\right)$. When $D_0 \gg \{ U_{dd},\epsilon_d,g \}$ is much larger than the microscopic model parameters, its precise value is immaterial\cite{hewson} and does not affect the low-energy Kondo physics. One then obtains,
\begin{eqnarray}\label{eq:lgret}
G^{(0)}_{\text{loc}} (\omega)&=&\mathcal{P}\int d\epsilon~
\frac{\rho(\epsilon)}{\omega-\epsilon}-i\pi \rho(\omega)\;, \nonumber
\\
&\overset{\omega \rightarrow 0}{=}&-\frac{\pi \ln
  \frac{\Lambda}{D_0}}{2|\omega| \ln^2 \big|\frac{\omega}{\Lambda}
  \big| } \left(\frac{2}{\pi}\ln \Big|\frac{\omega}{\Lambda} \Big| 
  {\rm{sgn}} (\omega)+i \right)\;,\nonumber \\
\end{eqnarray}
where $\mathcal{P}$ denotes the principal value, and the second line
is obtained analytically at low frequencies, as sketched in
Appendix~\ref{app:localG}. 
As demonstrated in the following, the model can now be solved
numerically exactly within NRG, exploiting a logarithmic
discretization of this hybridization function.  

We note that only interactions between the $d$-level and a \emph{single} (zero-mode) $\pi_p$ eigenmode were considered in Ref.~\onlinecite{cazalilla12}. However, this uncontrolled approximation to the full $H_{d\pi}$ can lead to a phase transition between doublet and triplet powerlaw Kondo states,\cite{cazalilla12} which we find to be an artefact of the approximation (all $\pi_p$ eigenmodes play and important role in the underlying Kondo physics).

In this paper we concentrate on the limit in which $U_{d\pi}$ and $J_H$ are both small compared with the induced Kondo coupling set by $J_K \propto g^2/U_{dd}$. For simplicity we now consider $U_{d\pi}=J_H=0$; although we have checked numerically that our results are robust to including small finite $d$-$\pi$ interactions. A detailed study of the full model as a function of  $U_{d\pi}$ and $J_H$ is technically complicated and beyond the scope of the current work. We postpone the full treatment of Eq.~(\ref{fullH}) to a future publication.\cite{graphene2}

On hybridization with the graphene conduction electrons, a low-energy
effective Kondo model can be derived using a Schrieffer-Wolff transformation,\cite{hewson} provided $g^2/U_{dd} \ll D_0$.  
Projecting onto the doublet (spin) manifold of $H_{d}$ by perturbatively eliminating virtual charge fluctuations to second order in the coupling $g$, one obtains 
\begin{eqnarray}\label{eq:Kondo}
H_K=H_{\pi}^0+J_K{\bf{S}}_d\cdot {\bf{s}} +V \sum_{\nu}
\tilde{\pi}_{\nu}^{\dagger} \tilde{\pi}_{\nu}^{\phantom{\dagger}}  
\end{eqnarray}
where $\textbf{S}_d$ is a spin-$\tfrac{1}{2}$ operator for the `impurity' d-level; and
${\bf{s}}=\frac{1}{2}\tilde{\pi}^{\dagger}_{\nu}\vec{\sigma}_{\nu
  \nu'}\tilde{\pi}^{\phantom{\dagger}}_{\nu'}$ is the local $\pi$ electron spin density at the defect site. The effective parameters of this Kondo model are,\cite{hewson}
\begin{eqnarray}\label{eq:SW}  
J_K&\simeq & 2 |g|^2 \left[
  \frac{1}{|\epsilon_d|}+\frac{1}{|U_{dd}+\epsilon_d|}
\right] \nonumber \\ 
V &\simeq & \frac{|g|^2}{2} \left[
  \frac{1}{|\epsilon_d|}-\frac{1}{|U_{dd}+\epsilon_d|}
\right] \;.
\end{eqnarray} 
It is important to note that the Kondo coupling $J_K>0$ is
\emph{antiferromagnetic}. Of course, the novel feature of this Kondo model
is the unusual density of states, Eq.~(\ref{eq:DOS}), whose modified
powerlaw behavior is shown to result in rich quantum impurity
physics. 

In real systems, one might expect that the pristine divergence in the local $\pi$-band density of states is cut off on some suitably low-energy scale. This might arise due to next-nearest neighbour hopping, or the presence of other vacancies and impurities. This experimentally relevant scenario is considered explicitly in Sec.~\ref{sec:genmodel}.


\subsection{Physical observables}\label{quantities}
In the following, we consider thermodynamic properties of
the above Anderson and Kondo models. 
In particular, we wish to calculate `impurity' (defect) contributions to the
entropy, $S_{\text{imp}}(T)$, and the magnetic 
susceptibility $\chi_{\text{imp}}(T)=\langle (\hat{S}_{\text{tot}}^z)^2
\rangle_{\text{imp}}/T$, as a function of temperature,
$T$. As usual,\cite{hewson} $\langle \hat{\Omega}
\rangle_{\text{imp}} = \langle \hat{\Omega}
\rangle -\langle \hat{\Omega}
\rangle_0$, with $\langle \hat{\Omega} \rangle_0$ denoting the thermal
average for the free system. 
Impurity contributions can thus be \emph{negative}, even though both 
$\langle \hat{\Omega}\rangle $ and $\langle \hat{\Omega} \rangle_0$ may separately be positive. 
The behavior of these quantities 
allows straightforward identification of the various fixed points, and evince
clear RG flow. Crossovers between these fixed points also provide
direct access to the underlying energy scales of the problem, such as
the Kondo temperature, $T_K$. 

In addition, we study dynamic quantities, focusing on the
energy-dependence of the t matrix. Since the t matrix controls 
the scanning tunneling spectroscopic response and resistivity measurements,
it is the key quantity needed to interpret certain experimental results. 
The rapid and universal evolution of the t matrix at energies on the order 
of the Kondo temperature could thus provide the `smoking gun' signature of 
Kondo physics in a graphene system with isolated vacancies. 

The t matrix describes scattering between diagonal states
$p$ and $p'$ of $H_{\pi}^0$ induced by the defect. 
It is generically given by,
\begin{equation}\label{eq:tmatrix}
\begin{split}
G_{p,p'}(\omega)=&G^{(0)}_{p,p}(\omega)\delta_{pp'}\\
&+G^{(0)}_{p,p}(\omega)\mathcal{T}_{p,p'}(\omega)
G^{(0)}_{p',p'}(\omega),
\end{split}
\end{equation}
where $\mathcal{T}_{p,p'}$ are components of the t matrix. Since hybridization between the $d$-level and the host
is local in space, $\mathcal{T}_{p,p'}(\omega)\equiv \frac{g_{p}g^{*}_{p'}}{g^2} T(\omega)$ and so one readily obtains,
\begin{equation}\label{eq:tmatrixloc}
G_{\text{loc}} (\omega)=G^{(0)}_{\text{loc}} (\omega)+\left [
  G^{(0)}_{\text{loc}} (\omega) \right ]^2 T(\omega) \; .
\end{equation}
For the full Anderson model, Eq.~(\ref{fullH}), the t matrix follows as
\begin{equation}\label{eq:tGimp}
T(\omega)=g^2 G_{dd}(\omega),
\end{equation}
where $G_{dd}(\omega)\equiv \langle\langle
d_{\nu}^{\phantom{\dagger}};d_{\nu}^{\dagger}\rangle\rangle_{\omega}$
is the $d$-level Green function, independent of spin $\nu$ in the absence of a magnetic field. 

At low energies in the Kondo regime, the same scattering should be
produced by the corresponding Kondo model, Eq.~(\ref{eq:Kondo}). The
t matrix is then expressed as,
\begin{equation}\label{eq:tmkondo}
T(\omega)=
T^{\textit{ps}}(\omega)+
T^{\textit{K}}(\omega)\;,
\end{equation}
where
\begin{subequations}\label{eq:tkondo}
\begin{align}
\label{eq:tkondops}
T^{\textit{ps}}(\omega)=&
\left (\frac{V}{1-V G^{(0)}_{\text{loc}} (\omega)}\right ) \\
\label{eq:tkondok}
T^{\textit{K}}(\omega)=&
\left ( \frac{J_K/2}{1-V G^{(0)}_{\text{loc}} (\omega)}
\right )^2 \langle\langle \hat{\gamma};\hat{\gamma}^{\dagger}\rangle\rangle_{\omega}
\;,
\end{align}
\end{subequations}
with $\hat{\gamma}=\hat{S}^z \tilde{\pi}_{\uparrow} + \hat{S}^{-}
\tilde{\pi}_{\downarrow}$ (here $\hat{S}^z$, $\hat{S}^{-}$ denote
operators for the `impurity' spin).\cite{akm:tqd2ch,akm:oddimp} $T^{\textit{ps}}(\omega)$ is the trivial contribution to scattering from a local potential, while $T^{\textit{K}}(\omega)$ describes the effect on scattering due to electron correlations, such as the Kondo effect. Below, we consider explicitly the spectrum of the t matrix, {\it i.e.},
\begin{equation}\label{eq:spec}
t(\omega)=-\frac{1}{\pi}\text{Im}~ T(\omega)\equiv -\frac{1}{\pi}\text{Im}\sum_p \mathcal{T}_{p,p}\;.
\end{equation}


\section{Kondo model}\label{sec:effkondo}

The Kondo model Eq.~(\ref{eq:Kondo}), with effective conduction electron density of states given by Eq.~(\ref{eq:DOS}), is a close relative of the so-called pseudogap Kondo
problem,\cite{WF,cassa,GBI,loganpt,lmapg,BGLlma,VF04,FV04,akm:qpiti,mvrev} where the
density of states is given generically by $\rho(\omega) \propto
|\omega|^r$. 
Such models have been studied extensively, especially in
the context of certain high-$T_c$ superconductors and magnetic
impurities in regular graphene,\cite{VFB10} which both realize the $r=1$
case. 
Vojta and Bulla also considered the case $-1<r<0$, describing a
spin-$\tfrac{1}{2}$ impurity coupled to a bath with diverging density of
states.\cite{VB02} Although their survey allowed the topology of the phase diagram to be identified, details of the various crossovers and
properties of the fixed points themselves were not established exactly. 
Further detailed study of the generalized powerlaw Kondo model for arbitrary
antiferromagnetic or ferromagnetic Kondo coupling will be considered
in a separate publication.\cite{MFBV}


\subsection{Analytical results}

\subsubsection{Generalized poor man's scaling}\label{sec:pms}

Before solving the Kondo model numerically exactly using NRG, we first consider perturbative scaling. A similar analysis was performed in
Ref.~\onlinecite{cazalilla12} for a Kondo model with a
related density of states (although the role of the scale $\Lambda$ was not considered).

In the spirit of Anderson's poor man's
scaling\cite{andersonpms,VB02,cazalilla12} one can derive flow equations 
for dimensionless couplings 
$j=\frac{J_K}{D \ln^2 \frac{D}{\Lambda}} \ln \frac{\Lambda}{D_0}$ and
$v=\frac{V}{D \ln^2 \frac{D}{\Lambda}} \ln \frac{\Lambda}{D_0}$. 
Two scaling functions arise for both the Kondo coupling $J_K$ and potential scattering $V$ (both of them diverging). Strong coupling scales $T_K$ and $T_P$ can be estimated and are shown to depend on $\Lambda$ as well as the bare couplings and $D_0$. This is in contrast to the situation in a regular metallic host, where potential scattering is irrelevant. 
Universal low-energy properties result from the flow of these couplings, and depend only on the low-energy scales $T_K$ or $T_P$. The high-energy band cutoff, $D_0$, is simply the bare energy scale where the microscopic parameters of the model are defined: we thus use it as our unit of energy in the following. We obtain,
\begin{eqnarray}
\frac{dj}{d \ln D}&=&\left(-1-\frac{2}{\ln
    \frac{D}{\Lambda}}\right)j-j^2 +\mathcal{O}(j^3)\;,\nonumber \\ 
\frac{dv}{d \ln
  D}&=&\left(-1-\frac{2}{\ln \frac{D}{\Lambda}}\right)v\;. 
\end{eqnarray}
Remarkably, the second equation describing scaling of the potential
scattering term is exact to all orders. 
The coefficient of the linear terms is the scaling dimension and here 
it depends upon the running scale $D$ and the bare scale 
$\Lambda$. Both scaling equations can be solved analytically to yield
the scale dependent couplings: 
\begin{eqnarray}\label{eq:scaling}
j(D)&=&
\frac{\rho(D)}{\rho(D_0)}\frac{j(D_0)}{1+\frac{j(D_0)}{\rho(D_0)}\int_{D_0}^D\frac{\rho(D')}{D'}dD'}
\;,\nonumber \\ 
v(D)&=&\frac{\rho(D)}{\rho(D_0)}v(D_0)\;. 
\end{eqnarray}
An immediate observation is that the strong coupling phases vanish as
$\Lambda \rightarrow D_0^+$ since then $j(D)$, $v(D)\rightarrow
0$. The density of states, Eq.~(\ref{eq:DOS}), reduces to a pair of
poles at $\pm D_0$ in this limit, and so there are no
low-energy bath degrees of freedom to which the impurity spin can
couple. However, in the generic case $\Lambda >D_0$, the running Kondo
coupling $j(D)$ and potential scattering $v(D)$ are renormalized
upward on successive reduction of the bandwidth/energy scale. 
The strong coupling scales $T_K$ and $T_P$ associated with the
divergence are scale invariants and can be identified approximately 
as the point where these couplings become of
order unity: {\it i.e.} $j\left(T_K\right)=1$ and
$v\left(T_P\right)=1$. Of course, such a perturbative approach breaks
down before this point, but the analysis does provide a useful
analytic estimate. 

To leading order, it suffices to consider the scaling dimension of the
flow equations. To this level of approximation, the flow of $j$ and
$v$ is identical, and so the resulting strong coupling scales have the
same functional dependence,
\begin{eqnarray}\label{eq:tk}
T_{K} \ln^2 \frac{T_{K}}{\Lambda}&=&\frac{J_K}{2}\ln \frac{\Lambda}{D_0}
\;,\nonumber \\
T_P \ln^2 \frac{T_P}{\Lambda}&=&\frac{V}{2}\ln \frac{\Lambda}{D_0}\;.
\end{eqnarray}
In the limit of small $J_K$ ($V$), and consequently small $T_K$ ($T_P$),
Eq.~(\ref{eq:tk}) can be inverted to yield 
\begin{eqnarray}\label{eq:pmsscales}
T_K &\approx & \frac{J_K}{2}\frac{\ln \frac{\Lambda}{D_0}}
{\ln^2 \frac{J_K}{\Lambda}} \; ,\nonumber \\
T_P &\approx & \frac{V}{2}\frac{\ln \frac{\Lambda}{D_0}}
{\ln^2 \frac{V}{\Lambda}}\;. 
\end{eqnarray}
The notable feature here is that the strong coupling scales are
proportional to the bare couplings $J_K$ and $V$. Of course, this is
in marked contrast to the standard metallic case, where $\rho(\omega)\equiv
\rho_0$ is finite and constant at low energies. An 
exponentially small Kondo temperature,\cite{hewson} $T_K\propto
\exp(-1/\rho_0 J_K)$, results in that case. Remarkably, strong coupling physics in the
defective graphene model should be expected at relatively high
temperatures.\cite{cazalilla12} Fundamentally, this is due to the enhanced conduction electron density of states at low energies. Indeed, experimental results on graphene systems with vacancies have shown that unusually high Kondo temperatures do result,\cite{fuhrer10} which was the original motivation\cite{cazalilla12} for Eq.~(\ref{eq:DOS}).

Below, we extract the exact dependence of $T_P$ on $V$ in the pure potential
scattering case ($J_K=0$). In Sec.~\ref{sec:tk} we go further and
extract $T_K$ and $T_P$ from exact non-perturbative NRG calculations
for the non-trivial case where $V\ne 0$ and $J_K\ne 0$.

The implication from the RG scaling equations, Eq.~(\ref{eq:scaling}), 
is that at high energies the physics is controlled by a `local moment' (LM) fixed point, describing a free impurity spin, decoupled from the bare host. For larger bare $J_K$ ($\gg V$), the Kondo coupling is renormalized faster as the temperature/energy scale is reduced, and hence the Kondo effect will dominate. The ground state is then described by a `symmetric strong coupling' (SSC) fixed point, and the impurity spin is screened by conduction electrons. The remaining Fermi liquid host degrees of freedom now also feel an additional phase shift (equivalent to a modified boundary condition at the defect site) due to the Kondo effect. In the case of a diverging free density of states, this results in an additional anomalous host contribution to physical properties, as shown below.   
Alternatively, if the potential scattering is initially very strong ($V \gg J_K$), then $V$ is renormalized faster and the ground state is described by the `asymmetric local moment' (ALM) fixed point. Although the impurity remains free, the renormalized potential scattering in the host becomes maximal and generates the same anomalous contribution to physical properties because the same phase-shift/boundary condition arises as in the Kondo-screened case. A critical point, AF-CR, separates the SSC and ALM phases.\cite{note:SCLM} The topology of the flow diagram is shown in Fig.~\ref{fig:pd}, and explicitly discussed in Sec.~\ref{sec:kondo} where the physical picture is confirmed by means of exact numerics. The underlying  topology of the phase diagram is thus equivalent to that studied in the \emph{pure} powerlaw case by Vojta and Bulla.\cite{VB02}


\subsubsection{Potential scatterer}\label{sec:potscat} 
The simplest limit of the Kondo model arises for $J_K=0$ in
Eq.~(\ref{eq:Kondo}). The model is then trivial in the sense that it
is non-interacting, and Kondo physics \emph{per se} is totally 
absent. However, as suggested by the flow equation,
Eq.~(\ref{eq:scaling}), this simple system has a strong coupling
ground state, characterized by potential scattering whose 
strength diverges below an emergent scale, $T_P$.  
Later, it will be shown that the low-energy behavior of the Kondo
model with $J_K > 0$ can also be understood in terms of the pure potential
scatterer throughout the ALM phase. We discuss briefly this limit now. 

The local Green function at the defect site is modified due to the
additional potential scattering for $V\ne 0$. It is expressed exactly
as,
\begin{eqnarray}\label{eq:psG}
\left [ G_{\text{loc}} (\omega)\right ]^{-1}=\left
[G^{(0)}_{\text{loc}} (\omega)\right ]^{-1} -V\;.
\end{eqnarray}
Rearranging this equation in the form of Eq.~(\ref{eq:tmatrixloc})
immediately yields the t matrix for the pure potential scatterer,
Eq.~(\ref{eq:tkondops}). 
The corresponding spectrum has the following asymptotic forms,
\begin{subequations}
\begin{align}
\nonumber t(\omega) \; \overset{|\omega|\ll |V|}{=}& \;
\left ( \frac{1}{2 \ln\frac{\Lambda}{D_0}} \right ) |\omega| -
\left ( \frac{~\text{sgn} [\omega]}{V \ln^2 \frac{\Lambda}{D_0}} \right ) |\omega|^2
\ln \left | \frac{\omega}{\Lambda}\right |\\ 
\label{eq:psspeclow}
 &\qquad\qquad\qquad\qquad\;\;+ \mathcal{O}(|\omega|^3) \;,\\
\nonumber \overset{|\omega|\gg |V|}{=}& \; t(\omega) =  \tfrac{1}{2} V^2 \ln
\frac{\Lambda}{D_0} \left ( \frac{1}{|\omega| \ln^2 |\omega/\Lambda|} \right )\\ 
\label{eq:psspechigh}
&\qquad\qquad\qquad\qquad\;\; +\mathcal{O}(1/|\omega|^2) \;. 
\end{align}
\end{subequations}
By demanding $\partial t(\omega) / \partial \omega =0$, we find that
the spectral peak for small $V$ arises when,
\begin{equation}\label{eq:pspeak}
V_P=-\omega_P \ln \left |\frac{\omega_P}{\Lambda} \right | / \ln
\frac{\Lambda}{D_0} \;.
\end{equation}
As $\omega_P\rightarrow 0$, the spectrum exhibits scaling in terms of
$\omega/\omega_P$, viz: 
\begin{subequations}
\label{eq:psscaling}
\begin{align}
\nonumber \frac{t(\omega)}{|\omega|} \; \overset{|\omega|\ll |\omega_P|}{=}& \;
\left ( \frac{1}{2 \ln \frac{\Lambda}{D_0}} \right )+
\left ( \frac{~\text{sgn} [\omega V]}{\ln \frac{\Lambda}{D_0}} \right )
\left |\frac{\omega}{\omega_P}\right | \\
\label{eq:psscalinglow}
&\qquad\qquad\qquad\qquad+ \mathcal{O}(|\omega/\omega_P|^2) \;,\\
\label{eq:psscalinghigh}
\overset{|\omega|\gg |\omega_P|}{=}& \; \left ( \frac{1}{2 \ln \frac{\Lambda}{D_0}}
\right )\left |\frac{\omega_P}{\omega} \right |^2+
\mathcal{O}(|\omega_P/\omega|^3) \;.
\end{align}
\end{subequations}
$\omega_P$ thus serves as a definition of the strong coupling scale,
$T_P\equiv  \omega_P$. The exact result in the limit $J_K=0$ is then, 
\begin{equation}\label{eq:psscale}
T_P\ln (T_P/\Lambda) = -|V|\ln \frac{\Lambda}{D_0} \;,
\end{equation}
which should be contrasted with the perturbative scaling result,
Eq.~(\ref{eq:tk}). The two definitions will be compared in
Sec.~(\ref{sec:tk}). 

The change in thermodynamic quantities due to the introduction of such potential scattering can also be obtained from the electron Green function, Eq.~(\ref{eq:tmatrix}), once the t matrix is known.\cite{hewson} At low temperatures, the behavior in the vicinity of the ALM fixed point is found to be,
\begin{subequations}
\label{eq:tdps}
\begin{align}
S_{\text{imp}}(T) =-\ln(2)&\left ( 1 +\frac{2}{\ln \frac{2T}{\Lambda}} \right )+\mathcal{O}(T^2)\;,\\
T\chi_{\text{imp}}(T) =\frac{1}{8}&\left ( 1-\frac{1}{\ln \frac{2T}{\Lambda}} \right )+\mathcal{O}(T^2)\;.
\end{align}
\end{subequations}
As confirmed by full NRG calculations below, this leading behavior arises on the lowest energy scales in the ALM phase also in the case of finite Kondo coupling, $J_K>0$.
 
It should be noted that these asymptotic results hold in the limit where the free conduction electron density of states is divergent, and described by Eq.~(\ref{eq:DOS}). In Sec.~\ref{sec:genmodel} we consider the case where this divergence is cut off at low energies.


\subsubsection{Resonant level}\label{sec:reslev}
In the Kondo screened SSC phase of the pseudogap Kondo model, it has
long been established\cite{loganpt,VF04,FV04} that the low-energy
physics is that of an \emph{effective} resonant level model,
\begin{equation}\label{eq:Hrl}
H_{\textit{RL}}=H_{\pi}^0 + \tilde{v} \sum_{\nu}
\left( \tilde{\pi}^{\dagger}_{\nu} d^{\phantom{\dagger}}_{\nu}
  +{\rm{H.c.}} \right) \;,
\end{equation}
which is just the non-interacting ($U=0$) Anderson
impurity model at particle-hole symmetry. Indeed, in the present case, 
we show below that the low-energy  
behavior of the SSC phase can again be understood in terms of the
resonant level. This motivates a brief analysis of the latter for the case
where $H_{\pi}^0$ is characterized by the local
density of states, Eq.~(\ref{eq:DOS}). 

The retarded resonant level Green function is simply,
\begin{equation}\label{eq:rlG}
G_{dd}(\omega)=\frac{1}{\omega+i0^{+}-\tilde{v}^2G_{\text{loc}}^{(0)}(\omega)}\;,
\end{equation}
where $G_{\text{loc}}^{(0)}(\omega)$ is given in
Eq.~(\ref{eq:lgret}). The corresponding spectrum is given 
asymptotically by, 
\begin{subequations}
\begin{align}
\nonumber t(\omega) \; \overset{|\omega|\ll |\tilde{v}|}{=}& \;
\left ( \frac{1}{2 \ln \frac{\Lambda}{D_0}} \right ) |\omega| -
\left ( \frac{1 }{\tilde{v}^2 \ln^2 \frac{\Lambda}{D_0}} \right ) |\omega|^3
\ln \left | \frac{\omega}{\Lambda}\right | \\
\label{eq:rlspeclow}
&\qquad\qquad\qquad\qquad\;\;+ \mathcal{O}(|\omega|^5) \;,\\
\nonumber\overset{|\omega|\gg |\tilde{v}|}{=}& \;   \tfrac{1}{2} \tilde{v}^4 \ln
\frac{\Lambda}{D_0} \left ( \frac{1}{|\omega|^3 \ln^2 |\omega/\Lambda|} \right ) \\
\label{eq:rlspechigh}
&\qquad\qquad\qquad\qquad\;\; +\mathcal{O}(1/|\omega|^5) \;. 
\end{align}
\end{subequations}
The position of the spectral peak at $\omega_K$ is straightforwardly obtained,
and the spectrum again obeys scaling as $\omega_K\rightarrow 0$ in
terms of $\omega/\omega_K$,
\begin{subequations}
\label{eq:rlscaling}
\begin{align}
\nonumber\frac{t(\omega)}{|\omega|} \; \overset{|\omega|\ll |\omega_K|}{=}& \;
\left ( \frac{1}{2 \ln \frac{\Lambda}{D_0}} \right )+
\left ( \frac{1}{\ln \frac{\Lambda}{D_0}} \right )
\left |\frac{\omega}{\omega_K}\right |^2 \\
\label{eq:rlscalinglow}
&\qquad\qquad\qquad\qquad+ \mathcal{O}(|\omega/\omega_K|^4) \;,\\
\label{eq:rlscalinghigh}
\overset{|\omega|\gg |\omega_K|}{=}& \; \left ( \frac{1}{2 \ln^2 \frac{\Lambda}{D_0}}
\right )\left |\frac{\omega_K}{\omega} \right |^4+
\mathcal{O}(|\omega_K/\omega|^6),
\end{align}
\end{subequations}
where $\omega_K^2\ln|\omega_K/\Lambda|=-\tilde{v}^2\ln(\Lambda/D_0)$. 

$H_{\textit{RL}}$ should be regarded as an \emph{effective}
low-energy model here, valid in the vicinity of the SSC fixed
point,\cite{VF04,FV04} and so the parameter $\tilde{v}$ is itself an
effective parameter. Thus, one cannot directly identify $\omega_K$ for the
resonant level model with $T_K$ for the full Kondo model with
$J_K>0$. However, the asymptotic scaling form of the spectrum, 
Eq.~(\ref{eq:rlscalinglow}), is expected to hold at low energies, as 
confirmed in the next section.

To lowest order, the RG scaling equations given in
Eq.~(\ref{eq:scaling}) are the same for the pure potential scatterer
(with $J_K=0$) and the Kondo model (with $V=0$). The strong coupling scale for the former is given by Eq.~(\ref{eq:psscale});  \emph{mutadis mutandis}, the leading dependence of the Kondo temperature in the latter should be,
\begin{equation}\label{eq:tkscale}
T_K\ln (T_K/\Lambda) = -|J_K|\ln \frac{\Lambda}{D_0} \;.
\end{equation}

On the lowest energy scales in the vicinity of the SSC fixed point, thermodynamics of the full Kondo model, Eq.~(\ref{eq:Kondo}), can be calculated from the effective resonant level model, Eq.~(\ref{eq:Hrl}).\cite{loganpt,VF04,FV04} The leading low-temperature behavior can again be extracted exactly from the electron Green function (or the corresponding resonant level t matrix), as shown explicitly for the entropy in Appendix \ref{app:logcorr}. As $T\rightarrow 0$ we find, 
\begin{subequations}
\label{eq:tdrl}
\begin{align}
S_{\text{imp}}(T) = -\ln(4)&\left ( 1 +\frac{1}{\ln \frac{2T}{\Lambda}} \right )+\mathcal{O}(T^2)\;,\\
T\chi_{\text{imp}}(T) =-\frac{1}{8}&\left ( 1+\frac{1}{\ln \frac{2T}{\Lambda}} \right )+\mathcal{O}(T^2)\;.
\end{align}
\end{subequations}
The low-temperature form of Eq.~(\ref{eq:tdrl}) is confirmed below by explicit NRG calculations for the full $S=\tfrac{1}{2}$ Kondo model in the SSC phase. Again we stress that these unusual results are obtained in the case where the conduction electron density of states has a pristine divergence described by Eq.~(\ref{eq:DOS}).


\subsection{Numerical results}\label{sec:kondo}

The physics of the Kondo model with finite $J_K$ and $V$ [Eq.~(\ref{eq:Kondo})] is obviously much more rich and subtle than the trivial limits considered above. Here, one generically expects two phases: for $J_K/|V|\gg 1$ a strong coupling SSC phase should result, while an ALM phase is stable for $|V|/J_K\gg 1$. These phases are separated by a quantum critical point (AF-CR) arising for a critical ratio $(J_K/V)_c=\mathcal{O}(1)$. 

\begin{figure}[t]
\includegraphics[width=0.48\textwidth]{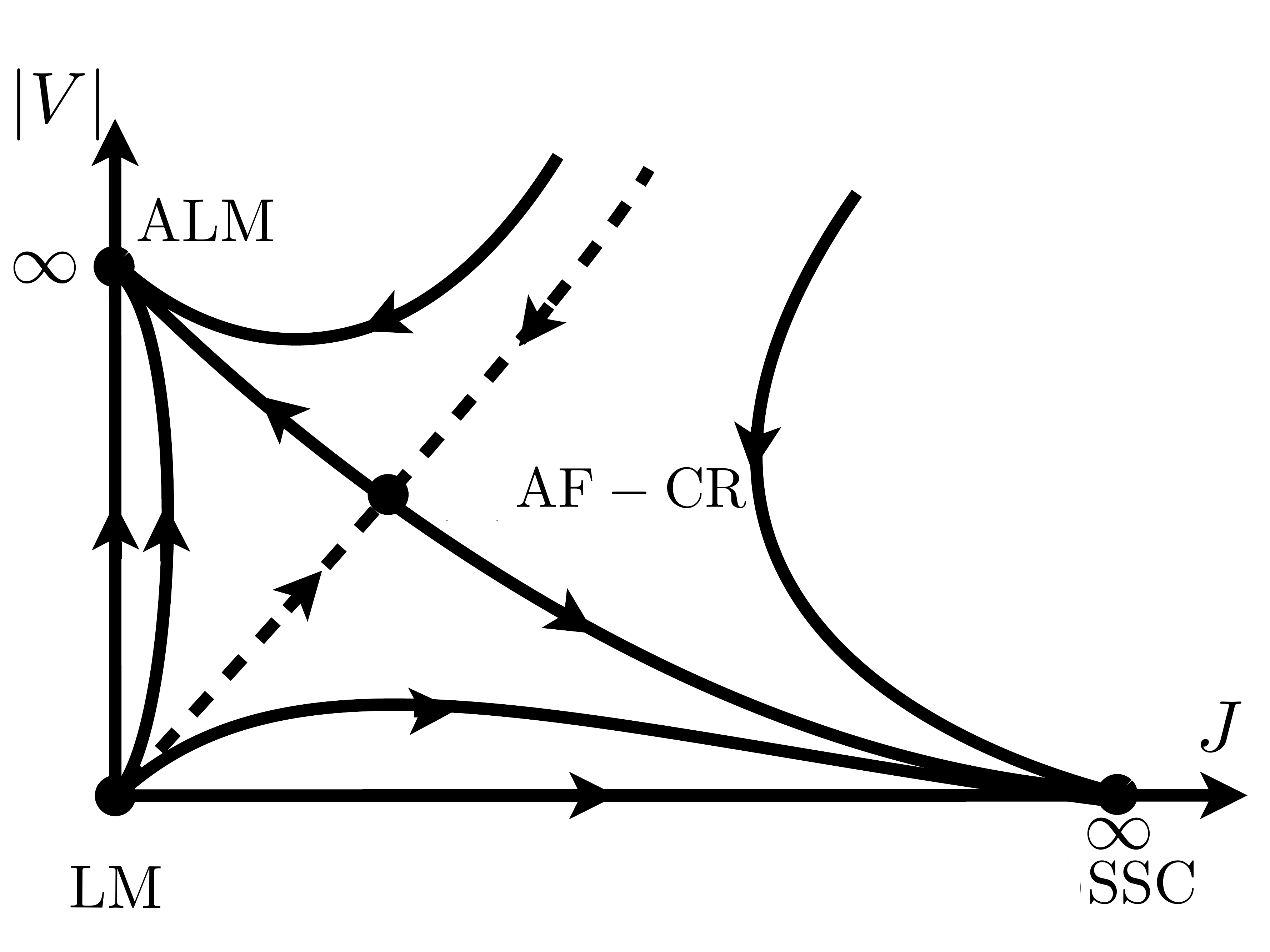}
\caption{Schematic phase diagram for the Kondo model with density of states given by Eq.~(\ref{eq:DOS}). $|V|$ denotes particle-hole symmetry breaking while $J_K$ is the Kondo coupling. Fixed points are denoted by circles, arrowed lines refer to effective RG flow, and the dashed line is the separatrix between SSC and ALM phases.}\label{fig:pd} 
\end{figure}

The full temperature-dependence of thermodynamic quantities can be calculated using NRG.\cite{nrg:rev} Their characteristic behavior at the various fixed points allows straightforward identification of the phases, and the entire phase diagram can thus be mapped out. We find that the topology of the phase diagram for antiferromagnetic $J_K>0$ is the same as for the case of the pure powerlaw density of states ($\rho(\omega)\sim |\omega|^r$ with $-1<r<0$) studied in Ref.~\onlinecite{VB02}. A schematic phase diagram is presented in Fig.~\ref{fig:pd} and discussed below.

At high energies/temperatures, the LM fixed point describes a free and unscreened impurity local moment. The limiting high-temperature entropy is thus $S_{\text{imp}}=\ln(2)$, while the magnetic susceptibility follows the Curie law, $T \chi_{\text{imp}}=1/4$. But the Kondo effect drives the system toward the SSC fixed point below an energy scale of order $T_K$ when $J_K/|V| > (J_K/|V|)_c$. For $T\ll T_K$ the entropy and susceptibility for the Kondo model are found from NRG to be given precisely by Eq.~(\ref{eq:tdrl}), obtained for the effective resonant level model. In particular, we note that the residual $T=0$ entropy $S_{\text{imp}}(0)=-\ln(4)$ is the smallest possible value, because the conduction electron density of states is characterized at low energies by the strongest possible divergence (up to logarithms) while remaining normalizable. 

By contrast, for $J_K/|V| < (J_K/|V|)_c$ the ALM fixed point is stable, and describes maximal particle-hole asymmetry. The crossover from LM to ALM physics occurs on the strong coupling scale, $T_P$. The free impurity local moment is then supplemented by an anomalous host contribution to give $T\rightarrow 0$ thermodynamics described by Eq.~(\ref{eq:tdrl}). The precise agreement on the lowest energy scales confirms the physical interpretation of the fixed point in terms of the pure potential scatterer discussed above.

On fine-tuning in the vicinity of the critical point $J_K/|V| \simeq (J_K/|V|)_c$, RG flow from the high-energy LM fixed point to the stable fixed point describing the ground state  occurs via the critical point, AF-CR. Two universal scales can thus be identified in this regime:  $T_c$ $(\approx T_K)$ sets the scale for onset of criticality associated with AF-CR, while $T^*$ $(\propto |V-V_c|)$ characterizes the ultimate crossover to either SSC or ALM fixed points (depending on the sign of $V-V_c$). The distinctive asymptotic thermodynamic properties of each fixed point are summarized in Table~\ref{tab:s12}.

\begin{figure}[t]
\includegraphics[width=0.49\textwidth]{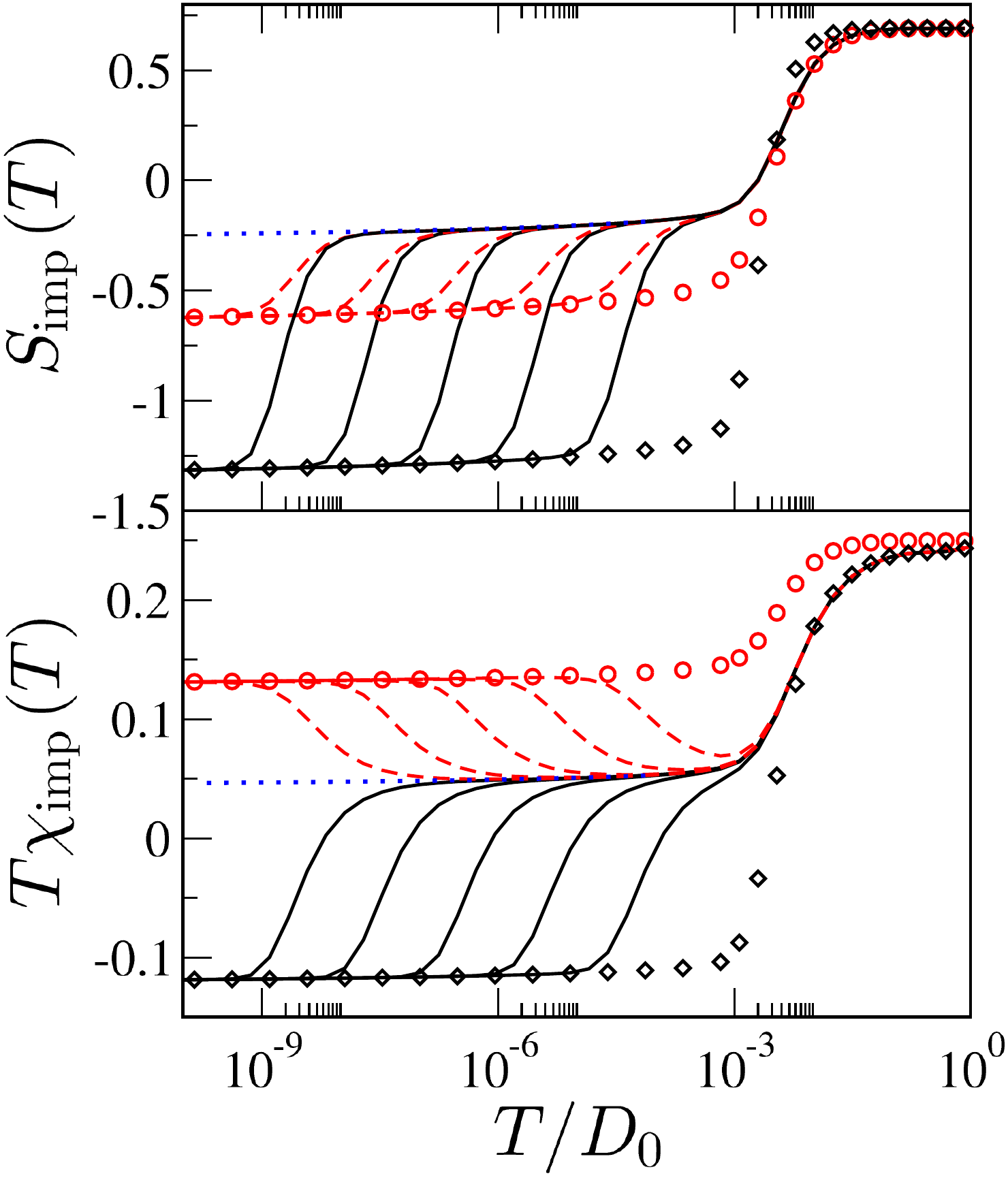}
\caption{Thermodynamics across the quantum phase transition. Entropy $S_{\text{imp}}(T)$ (upper panel) and magnetic susceptibility $T\chi_{\text{imp}}(T)$ (lower
panel) plotted vs $T/D_0$ for fixed $\Lambda/D_0=1.5$, $J_K=0.1 D_0$ 
and tuning $V \rightarrow V_c$. 
Shown for $V=V_c\pm 10^{-n} T_K^{V=0}$ with $n=0,1,2,3,4,5$ in
order of decreasing $T^*$ ($\propto |V-Vc|$). Solid lines for the SSC
phase; dashed lines for the ALM phase. The critical point itself is shown as the dotted lines. For comparison: diamond points for $J_K/D_0=0.1$, $V/D_0=0$ and circle points for $J_K/D_0=0$, $V/D_0=0.1$.}\label{fig:thermo}
\end{figure}

\begin{table}[t]
\caption{Properties of fixed points}
 \begin{tabular}{ l || c | c }
    \hline
    \phantom{1} & $\lim_{T\to 0}S_{\rm{imp}}$ & $\lim_{T\to 0} T \chi_{\rm{imp}}$ \\ \hline
    ALM & $-\ln 2$ & 1/8 \\ \hline
    LM & $\ln 2$ & 1/4 \\ \hline 
    AF-CR & $-\ln 4/3$ & 1/24 \\ \hline        
    SSC & $-\ln 4$ & -1/8 \\
    \hline
  \end{tabular}\label{tab:s12}
\end{table}

Remarkably, the impurity entropy at ALM, AF-CR, and 
SSC fixed points is negative. We note, however, that all thermodynamic quantities flow to their respective zero-temperature values logarithmically slowly. This is a direct consequence of the logarithmic energy dependence in the density of states, see Eq.~\eqref{eq:DOS}.

Fig.~\ref{fig:thermo} shows the full crossovers in the entropy and susceptibility calculated numerically exactly using NRG. In both cases, diamond points show the direct crossover from LM to SSC arising for $J_K/D_0=0.1$ but $V/D_0=0$; while circle points show the direct crossover from LM to ALM when $V/D_0=0.1$ but $J_K/D_0=0$. Thermodynamics are also shown on tuning to the quantum critical point (dotted line). Approach from the SSC phase (solid lines) and the ALM phase (dashed lines) exhibit two characteristic scales, $T_c$ and $T^*$, as above.

\begin{figure}[t]
\includegraphics[width=0.49\textwidth]{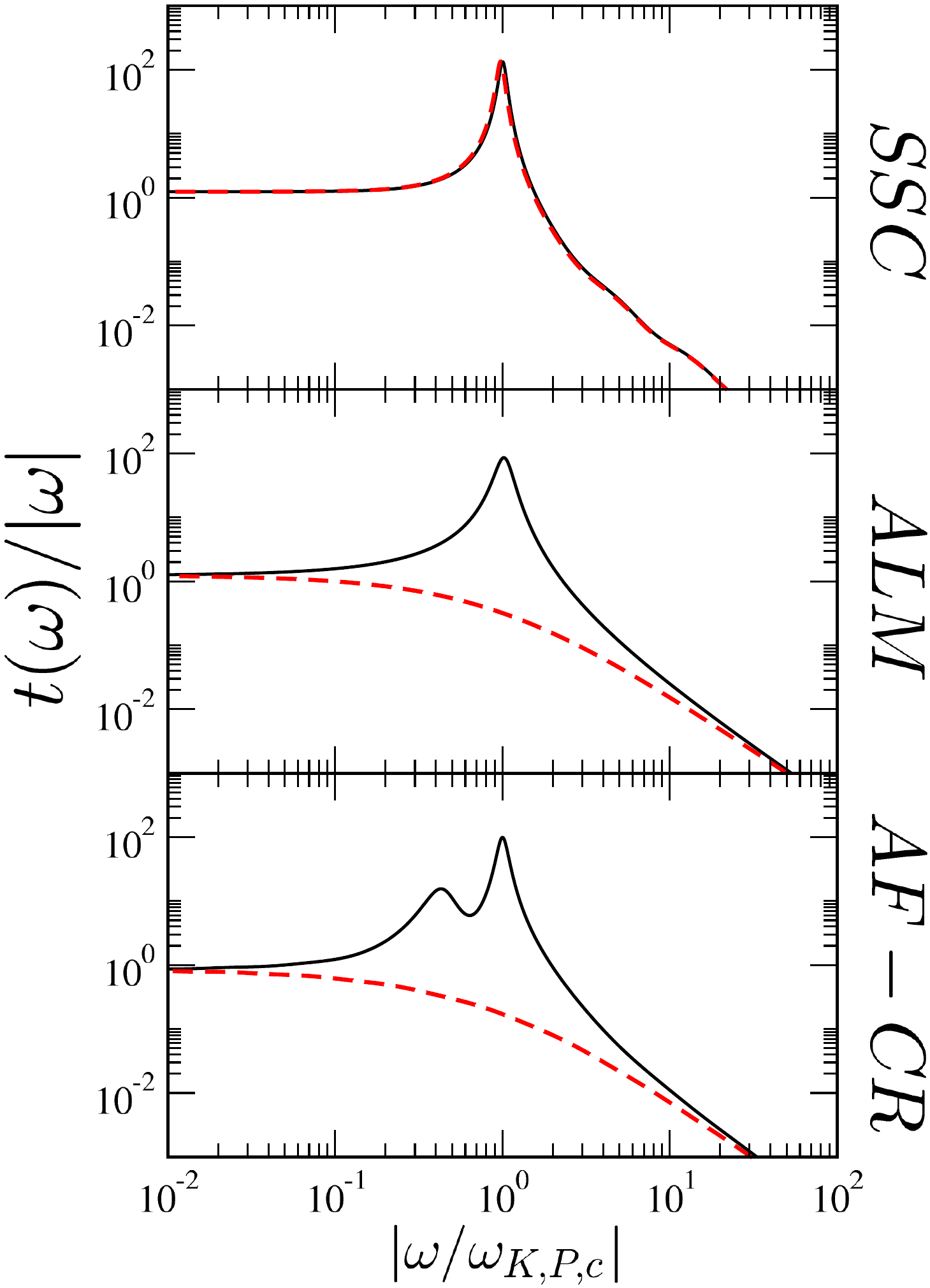}
\caption{Spectral function $t(\omega)/|\omega|$ at $T=0$. $\omega>0$ and $\omega<0$ plotted as solid and dashed lines.
Top panel: vs $|\omega/\omega_K|$ in the SSC phase
($J_K=10^{-6}D_0$ and $V/J_K=0.01$). 
Middle panel: vs $|\omega/\omega_P|$ in the ALM phase ($V=10^{-6}D_0$ and
$J_K/V=0.01$). Lower panel: vs $|\omega/\omega_c|$ at the critical
point ($J_K=10^{-6}D_0$ and $V/J_K=(V/J_K)_c$). $\Lambda/D_0=1.5$ is used throughout. Asymptotes discussed in the text.}\label{fig:tmatrix}
\end{figure}

Dynamical quantities, such as the $T=0$ scattering t matrix, similarly evince the rich RG structure of the problem. The NRG method also allows calculation of such dynamics,\cite{nrg:rev} which have been shown to be numerically exact in cases where exact results are known.\cite{akm:exactNFL,akm:finiteT} In Fig.~\ref{fig:tmatrix} (upper panel) we plot the scaling spectra $t(\omega)/|\omega|$ vs $|\omega/\omega_K|$ for a system deep in the SSC phase. Its asymptotic behavior at both high and low energies is found to be described by Eq.~(\ref{eq:rlscaling}).
In the ALM phase (Fig.~\ref{fig:tmatrix}, center panel), the scaling spectrum $t(\omega)/|\omega|$ vs $|\omega/\omega_P|$ is asymptotically described by Eq.~(\ref{eq:psscaling}). In particular, we note the more gentle linear approach to the Fermi level, and the inherent asymmetry of the spectrum, arising due to the relevance of particle-hole symmetry-breaking.
The lower panel of Fig.~\ref{fig:tmatrix} shows a critical spectrum, plotted as $t(\omega)/|\omega|$ vs $|\omega/\omega_c|$. The resonance around $\omega=\text{sgn} [\omega V] |\omega_c|$ is split, with the peaks separated by $\sim |V-J_K|$. 
At low energies $|\omega| \ll |\omega_c|$ the spectrum $t(\omega)/|\omega|\sim a+ b~\text{sgn} [\omega V] |\omega/\omega_c|$ has a leading linear dependence.

\subsection{Evolution of $T_K$ and $T_P$}\label{sec:tk}

Solution of the generalized poor man's scaling equations gives an estimate for the strong coupling scales $T_K$ and $T_P$, Eq.~(\ref{eq:tk}). In the trivial potential scattering limit considered explicitly in Sec.~\ref{sec:potscat}, the scale $T_P$ can be obtained exactly, Eq.~(\ref{eq:psscale}). Up to logarithms, one thus expects in either case the same linear dependence of the strong coupling scale $T_P\sim V$ or $T_K \sim J_K$.

\begin{figure}[t]
\includegraphics[width=0.45\textwidth]{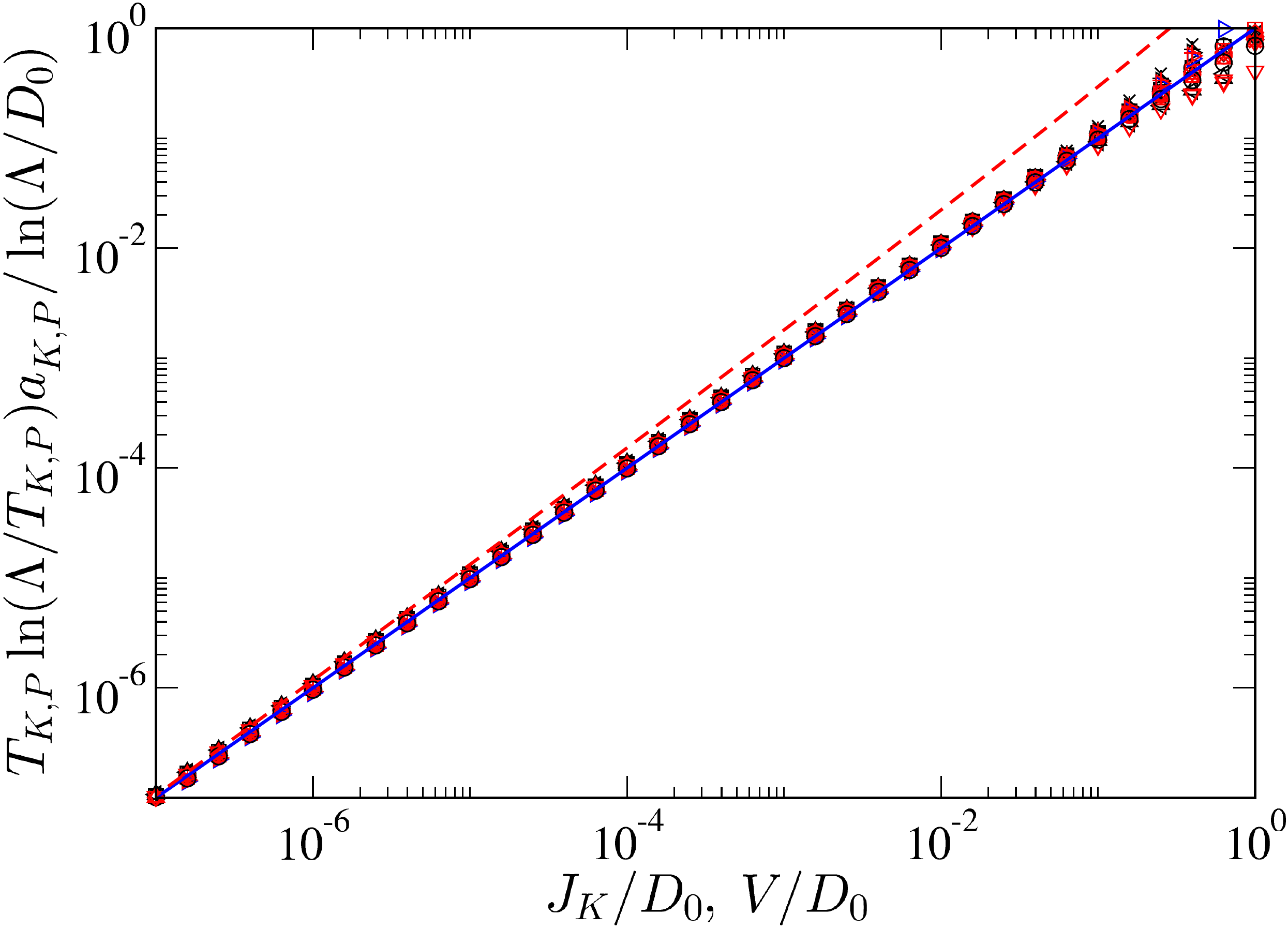}
\caption{$T_K$ and $T_P$ extracted from NRG calculations for various values of
$J_K$ and $V$ deep in the SSC phase ($J_K/V=2$) and the ALM phase
($J_K/V=1/2$) for $\Lambda/D_0=100, 10, 1.1,
1.01, 1.001$. All data plotted as $T_{K,P} \ln(\Lambda/T_{K,P})
a_{K,P}/\ln(\Lambda/D_0)$ vs $J_K/D_0$ or $V/D_0$ [the constant $a_{K,P} =
\mathcal{O}(1)$ depends only on the phase (SSC or ALM)]. Data collapse to the analytic result
[Eq.~(\ref{eq:psscale}), solid line], is found at small $J_K$, $V$. For comparison, Eq.~(\ref{eq:tk}) for $\Lambda/D_0=10$ is shown as the dashed line.}\label{fig:tk}
\end{figure}

Here we calculate $T_P$ and $T_K$ numerically exactly using NRG,\cite{note:tkcalc} varying $V$, $J_K$ and the cutoff $\Lambda$. In Fig.~\ref{fig:tk} we plot $T_K \ln(\Lambda/T_K)/\ln(\Lambda/D_0)$ vs $J_K/D_0$ for systems in the SSC phase, and $T_P \ln(\Lambda/T_P)/\ln(\Lambda/D_0)$ vs $V/D_0$ for systems in the ALM phase. The excellent agreement, especially for small $J_K$ or $V$, confirms Eqs.~(\ref{eq:tkscale}) and (\ref{eq:psscale}) (solid line). For comparison, the result of Eq.~(\ref{eq:tk}) is shown as the dashed line.

The key point is that the Kondo temperature is typically rather large in these systems.

\section{Anderson model}\label{sec:anderson}

\subsection{Accessibility of the ALM phases}
In Sec.~\ref{sec:effkondo} we discussed the rich phase diagram of
the Kondo model with conduction electron density of states given by Eq.~(\ref{eq:DOS}). 
This model was derived from an Anderson model (which also allows for charge fluctuations), by projecting onto the `impurity' spin-manifold using a Schrieffer-Wolff transformation.  Within this leading-order perturbative treatment, Eq.~\eqref{eq:SW} indicates that the maximum ratio of effective potential scattering, $V$, and effective Kondo exchange coupling, $J_K$, is given by
 \begin{eqnarray}\label{eq:swratio}
\bigg| \frac{V}{J_K} \bigg| \leq \frac{1}{4}\;.
 \end{eqnarray}
The natural question is then: which of the strong coupling phases of
a Kondo model can actually be accessed within the bare Anderson model. 
In order to answer this, we study the exact position of the phase boundary of the 
effective Kondo model and examine directly the full Anderson model, Eq.~(\ref{fullH}).

\begin{figure}[t]
\includegraphics[width=0.49\textwidth]{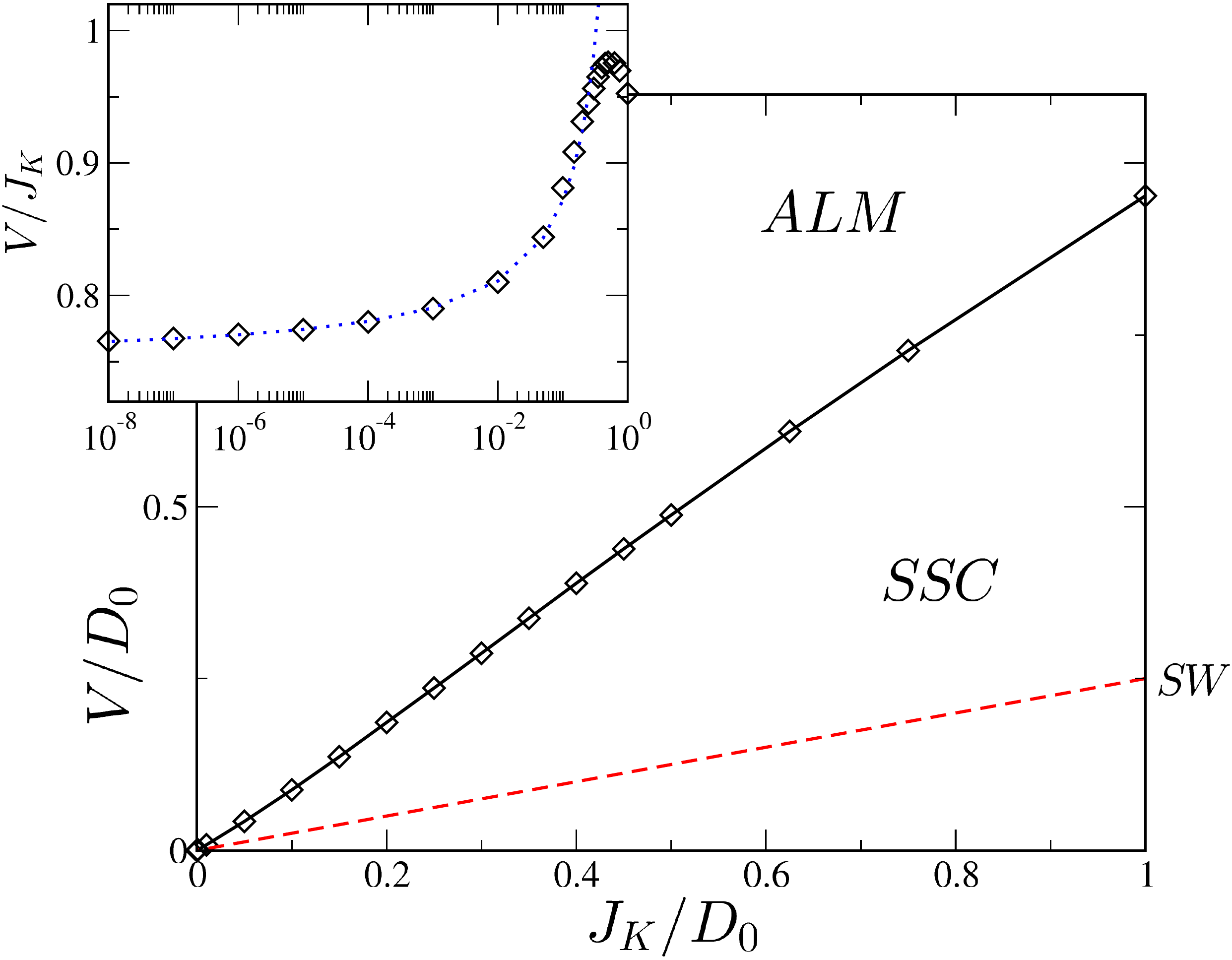}
\caption{Phase diagram in the $J_K-V$ plane for $\Lambda/D_0=1.5$, 
indicating the critical separatrix 
between SSC and ALM phases. At small
$J_K$, the critical ratio $|V/J_K|_c$ is given by Eq.~(\ref{eq:smallVpd}) 
[see inset, dotted line]. The regime of validity of the effective Kondo model is given by
the Schrieffer-Wolff asymptote $|V/J_K|_{\text{max}}=1/4$, shown as the
dashed line.}\label{fig:sc}
\end{figure}

Fig.~\ref{fig:sc} shows the phase diagram of the Kondo model obtained using NRG. The inset shows the asymptotic small-$J_K$ behavior, which is found to follow
\begin{equation}\label{eq:smallVpd}
\left | \frac{V}{J_K} \right |_c = a+b\ln(J_K/\Lambda),
\end{equation}
with $a=3/4$ and $b\approx -0.3$ for $\Lambda/D_0=1.5$ (see dotted line, inset).

For comparison, we also plot the Schrieffer-Wolff result, Eq.~(\ref{eq:swratio}), as the dashed line in Fig.~\ref{fig:sc}. The obvious conclusion is that the effective potential scattering, $V$, derived from the bare Anderson model is never large enough to access the ALM phase. A numerical survey of the parameter space of Eq.~(\ref{fullH}) supports this result, and suggests that the Kondo effect is always operative at the lowest energy scales. The ground state is thus described by the SSC fixed point when the density of states is given by  Eq.~\eqref{eq:DOS}.


\section{Accessibility of powerlaw Kondo physics in graphene}\label{sec:genmodel}

In the previous sections, we used a simplified conduction electron density of states, Eq.~(\ref{eq:DOS}), obtained\cite{cazalilla12} within a nearest-neighbour tight-binding approximation. Going beyond this approximation, one might also expect the $d$-level to hybridize weakly with more distant $\pi$ orbitals (albeit with extremely small hopping amplitude, given the local nature of the structural corrugations around the defect). Indeed, weak inter-sublattice coupling between next-nearest-neighbours may also play a role, leading to a sharp \emph{resonance} rather than a true logarithmic divergence in the density of states.

To simulate these effects heuristically, we introduce an effective density of states,
\begin{eqnarray}\label{modifieddos}
\rho (\omega)=\frac{\tfrac{1}{N}\theta \left(D_0-|\omega|\right)}{(|\omega|+X)
  \ln^2 \tfrac{1}{\Lambda}(|\omega|+X)}~\;,
\end{eqnarray}
defined inside a band of half-width $D_0$, which is normalizable for $\Lambda > D_0$, whence $N=2[\ln^{-1}(X/\Lambda) - \ln^{-1}\tfrac{1}{\Lambda}(D_0+X)]$. As such, it recovers Eq.~(\ref{eq:DOS}) in the limit $X\rightarrow 0$. For small finite $X$, the divergence is cut off on the scale of $|\omega| \sim X$, and the density of states becomes flat for $|\omega| \ll X$. Although $X$ might be very small in practice, it is expected to have an effect on the lowest energy/temperature scales.

\begin{figure}[t]
\includegraphics[width=0.49\textwidth]{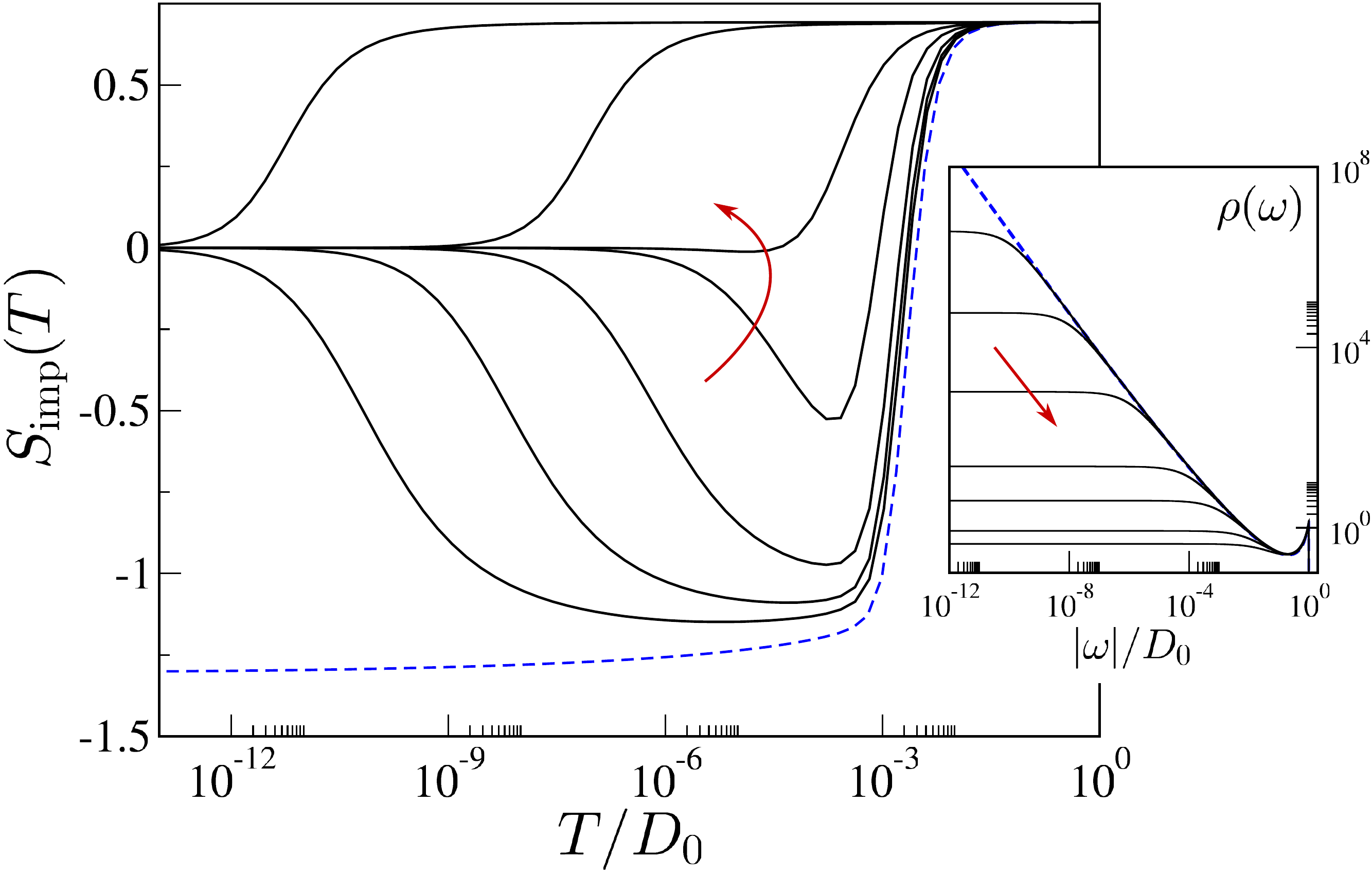}
\caption{Entropy for the Kondo model with density of states given by Eq.~(\ref{modifieddos}). Plotted for $\Lambda/D_0 = 1.5$, $J_K/D_0=0.1$,
$V/D_0=0$, increasing $X/D_0$ in the direction of the arrow from 0 (blue dashed 
line) to $X/D_0=10^{-10}, 10^{-8}, 10^{-6}, 10^{-4}, 10^{-3}, 10^{-2}, 10^{-1.5}$. Inset shows
the corresponding density of states. 
}\label{fig:Xvar}
\end{figure}

Provided $X$ is small (as might be expected physically), the results of the previous sections should however hold in the temperature/energy regime $X\ll T$. We explore this scenario in Fig.~\ref{fig:Xvar}, where we consider explicitly a spin-$\tfrac{1}{2}$ Kondo model with bath density of states given by Eq.~(\ref{modifieddos}). We take fixed $J_K/D_0=0.1$, $V/D_0=0$, $\Lambda/D_0=1.5$ and increase $X/D_0$ in the direction of the arrow (the blue dashed line corresponds to $X=0$). 

For $X\ll T_K^{X\rightarrow 0}$, there is an extended temperature regime $X\ll T \ll T_K^{X\rightarrow 0}$ where the system flows near to the SSC fixed point [and thus approaches the limiting entropy $-\ln(4)$]. However, on the lowest energy scales $T\ll X$, RG flow is ultimately to the regular Kondo SC fixed point,\cite{hewson} with all residual entropy quenched, $S_{\text{imp}}(T\rightarrow 0)=0$.

Even when the divergence is cut off on the scale of $X$, the low-energy density of states is still greatly enhanced when $X$ is small. Importantly, this leads to a large Kondo temperature. As shown in Fig.~\ref{fig:Xvar}, the Kondo temperature diminishes very rapidly as $X$ is increased, and conduction electron spectral weight is moved away from the Fermi level. Experiments on vacancies in graphene have in fact found surprisingly high Kondo temperatures,\cite{fuhrer10} which suggests that $X$ is in practice rather small. As a consequence, the distinctive signatures of the powerlaw Kondo effect (as considered in the previous sections), should appear in an intermediate temperature window.

\section{Conclusions}\label{sec:conclusions}

In this paper we studied the physics of a quantum impurity model with diverging hybridization, as might be obtained at the site of a reconstructed vacancy in graphene.\cite{cazalilla12} A dangling orbital of the graphene $\sigma$ band localized at the defect can hybridize with $\pi$ conduction electrons. We consider the situation where onsite interactions of the localized $\sigma$-level are strong, but $\sigma$-$\pi$ interactions are weak. A combination of analytical and numerical techniques were employed to obtain a rather complete understanding of the model and its properties. 

When charge fluctuations are frozen out, the Anderson model can be more simply understood in terms of an effective spin-$\tfrac{1}{2}$ `impurity' exchange-coupled to $\pi$-band conduction electrons of the graphene host. This Kondo model is unusual due to the diverging hybridization, and number of distinctive physical properties result. Indeed, we find a rich phase diagram, arising because particle-hole symmetry-breaking is relevant in this system, unlike the situation in standard metals. Strong coupling phases associated with large renormalized Kondo coupling (SSC) or potential scattering (ALM) are thus supported. 

Interestingly, the graphene conduction electrons in \emph{both} cases feel a $\pi/2$ phase shift as $T\rightarrow 0$: in the SSC phase by spin-singlet formation with the impurity, and in the ALM phase by the renormalized potential scattering. The localized bath orbital at the defect is thus projected out in both phases, giving rise to anomalous contributions to ground state thermodynamic properties. 
For example, even in the more conventional Kondo SSC phase, the residual impurity contribution to entropy assumes the minimum possible value, $S^{\textit{SSC}}_{\text{imp}}(T\rightarrow 0)=-k_B\ln(4)$ --- despite the impurity spin itself being quenched by the Kondo effect. We also find a strongly enhanced \emph{linear} scaling of the Kondo temperature with coupling strength in this phase, $T_K \sim J_K$ (up to logarithms). By contrast, the ALM phase is characterized by a large renormalized potential scattering, which suppresses the Kondo effect. The asymptotically-free impurity local moment is however similarly supplemented by the anomalous bath contribution, yielding now $S^{\textit{ALM}}_{\text{imp}}(T\rightarrow 0)=-k_B\ln(2)$. 

The lowest-energy physics in the vicinity of the SSC and ALM stable fixed points was understood analytically in terms of effective resonant level and pure potential scattering models; and the physical picture confirmed by means of exact numerics. The phases are separated by an unstable quantum critical point, which was also studied in detail.

However, direct analysis of the full Anderson model reveals that the phase transition separating the two strong coupling phases cannot in practice be accessed, because the effective Kondo coupling and effective potential scattering are \emph{slaved}. We stress this important cautionary caveat when dealing with Kondo models in general: the parameters of the effective Kondo model are not independent, since they depend on the same microscopic parameters of the underlying Anderson model. In the present case, the Kondo effect is thus \emph{always} operative on the lowest energy scales (of course, additional potential scattering from other sources could manifest at higher energies $|\omega| > \Lambda$, not considered here). 

Finally, we comment on the accessibility of the above physics in real systems. 
The pristine divergence of the graphene $\pi$-band conduction electron density of states suggested in Ref.~\onlinecite{cazalilla12} might more realistically be cut off at low energies. We showed however that exotic physical behavior controlled by the modified powerlaw Kondo effect might still be accessible in an intermediate temperature/energy window. Only on the lowest energy scales does conventional metallic Kondo physics dominate. 
Indeed, experiments on graphene samples with vacancies have revealed unusually large Kondo temperatures,\cite{fuhrer10} consistent with the above picture. 

One assumption employed in this work was to neglect capacitive and Hund's rule interactions between the local $\sigma$ level at the defect and $\pi$-band conduction electrons. Although we have checked that the physics discussed in this paper is robust to including small $\sigma$-$\pi$ interactions of this type, preliminary results suggest that new phases and physics can also be accessed when these interactions are stronger. This will be the subject of a future publication. 
\\


\subsubsection*{Acknowledgements}
We acknowledge useful discussions with Ralf Bulla, Matthias Vojta and
Martin Galpin. This work was supported by the DFG under FR 2627/3-1
(LF), SFB 608 (AKM,LF) and FOR 960 (AKM); and by the EPSRC through EP/1032487/1 (AKM).


\appendix
\section{Local Green function}\label{app:localG}
The derivation of the local Green function at low frequencies is straightforward but tedious. In the following we sketch it: while the imaginary part follows trivially from the density of states the real part has to be evaluated from Kramers-Kronig relation
\begin{eqnarray}
G^{(0)}_{\text{loc}} (\omega)&=&\mathcal{P}\int d\epsilon~
\frac{\rho(\epsilon)}{\omega-\epsilon}-i\pi \rho(\omega)\;.
\end{eqnarray}
The real part can be brought into the form
\begin{eqnarray}
\Re G^{(0)}_{\text{loc}} (\omega)=\ln \frac{\Lambda}{D_0}\frac{{\rm{sgn}}(\omega)}{|\omega|}\mathcal{P}\int_0^{\frac{D_0}{|\omega|}}\frac{dE}{(1-E^2)E \ln^2 \frac{E|\omega|}{\Lambda}}.\nonumber \\ 
\end{eqnarray}
After an integration by parts one arrives at
\begin{eqnarray}
\frac{{\rm{sgn}}(\omega)|\omega|}{\ln \frac{\Lambda}{D_0}}\Re G^{(0)}_{\text{loc}} (\omega)&=& \lim_{\delta \to 0} \left[ -\frac{1}{\ln \frac{E |\omega|}{\Lambda}(1-E^2)} \right]_{0}^{1-\delta} \nonumber \\ &+& \lim_{\delta \to 0} \left[ -\frac{1}{\ln \frac{E |\omega|}{\Lambda}(1-E^2)} \right]_{1+\delta}^{\frac{D_0}{|\omega|}} \nonumber \\ &+& \lim_{\delta \to 0} \int_0^{1-\delta} \frac{2 E dE}{(1-E^2)^2 \ln \frac{E |\omega|}{\Lambda}} \nonumber \\ &+& \lim_{\delta \to 0} \int_{1+\delta}^{\frac{D_0}{|\omega|}} \frac{2 E dE}{(1-E^2)^2 \ln \frac{E |\omega|}{\Lambda}} \;.\nonumber \\
\end{eqnarray}
At this point it is important to notice that in the limit $|\omega| \ll \Lambda$ we can rewrite the above expression to leading order
\begin{eqnarray}
\frac{{\rm{sgn}}(\omega)|\omega|}{\ln \frac{\Lambda}{D_0}}\Re G^{(0)}_{\text{loc}} (\omega)&\approx& \lim_{\delta \to 0} \left[ -\frac{1}{\ln \frac{E |\omega|}{\Lambda}(1-E^2)} \right]_{0}^{1-\delta} \nonumber \\ &+& \lim_{\delta \to 0} \left[ -\frac{1}{\ln \frac{E |\omega|}{\Lambda}(1-E^2)} \right]_{1+\delta}^{\frac{D_0}{|\omega|}} \nonumber \\ &+& \lim_{\delta \to 0} \frac{1}{\ln \frac{|\omega|}{\Lambda}} \int_0^{1-\delta} \frac{2 E dE}{(1-E^2)^2 } \nonumber \\ &+& \lim_{\delta \to 0}\frac{1}{\ln \frac{|\omega|}{\Lambda}}  \int_{1+\delta}^{\frac{D_0}{|\omega|}} \frac{2 E dE}{(1-E^2)^2} \;. \nonumber \\
\end{eqnarray}
This expression directly leads to Eq.~\eqref{eq:lgret}.

\section{Logarithmic corrections}\label{app:logcorr}

Here we obtain analytically the first logarithmic correction to
the impurity entropy, from the free energy of an effective resonant level
model. The general formula to calculate the free energy of a local
level is given by 
\begin{eqnarray}
\mathcal{F}=-T \sum_{\sigma} \sum_{\omega_n} \ln \left( -
  G^{-1}_\sigma \left(\omega_n \right) \right)e^{i\omega_n0^+} 
\end{eqnarray}
Using residual calculus this can be converted into a line integral
along the branch cut along the real axis given by 
\begin{eqnarray}
\mathcal{F}=- \sum_\sigma\int_{-\infty}^\infty \frac{dz}{\pi} n_F(z)
e^{z0^+} {\rm{Im}}\,\ln \left( -G^{-1}_{r,\sigma} \left(z
  \right)\right)  
\end{eqnarray}

where the subscript $r$ refers to the retarded Green function. 
Consequently, we obtain the entropy as 
\begin{eqnarray}
S=-\frac{\partial \mathcal{F}}{\partial T}=\int_{\infty}^\infty
\frac{dz}{\pi} \frac{2}{\cosh^2 z} {\rm{Im}}\,\ln \left(
  -G^{-1}_{r,\sigma} \left(z 2 T\right)\right) \;, \nonumber \\ 
\end{eqnarray}
where the factor of two is due to the spin summation. 
The low-energy properties will be entirely dominated by the
self-energy, which is of the form  
\begin{eqnarray}
\Sigma(z) \approx \tilde{v}^2 \frac{\pi \ln \frac{\Lambda}{D_0}}{2|z| \ln^2
  \big|\frac{z}{\Lambda} \big| } \left(\ln \Big|\frac{z}{\Lambda}
  \Big| \frac{2}{\pi} {\rm{sgn}} (z)+i \right)\;. 
\end{eqnarray}
It is then straightforward to derive an expression for the impurity entropy.  It follows as, 
\begin{eqnarray}
S=-\int_{0}^\infty \frac{d z}{\pi} \frac{2z}{\cosh^2 z} \left( \pi+2
  \arctan \frac{\pi}{2\ln \big|\frac{2T z}{\Lambda} \big| }  \right) 
\end{eqnarray}
Realizing the low-temperature limit, this can be approximated as,
\begin{eqnarray}
S\approx-\int_{0}^\infty d z \frac{2z}{\cosh^2 z} \left( 1+
  \frac{1}{\ln \big|\frac{2T z}{\Lambda} \big|}  \right) 
\end{eqnarray}
which, to leading order, yields
\begin{eqnarray}
S\approx- \ln 4 \left( 1+ \frac{1}{\ln \big|\frac{2T }{\Lambda} \big|}  \right)\;.
\end{eqnarray}




\begin{thebibliography}{43}
\bibitem[{not({\natexlab{b}})}]{note:tkcalc}
\bibinfo{note}{The values of $T_K$ and $T_P$ are determined in practice from
  NRG as the point where the impurity contribution to the entropy is halfway
  between the high-temperature LM fixed point and low-temperature SSC or ALM
  fixed point values:
  $S_{\text{imp}}(T_{K,P})=\tfrac{1}{2}[S_{\text{imp}}(\infty)-S_{\text{imp}}(%
0)]$.}
\end{thebibliography}
\end{document}